\documentclass[amsmath,amssymb,aps,prd,reprint,groupedaddress]{revtex4-2}
\usepackage{graphicx}% Include figure files
\usepackage{dcolumn}% Align table columns on decimal point
\usepackage{bm}% bold math
\usepackage{braket}
\usepackage{xcolor}

\begin{document}

%\preprint{}

\title{$q\bar{q}$ scattering phase shift in the $\pi^0$ channel and ${\pi}^0$ meson spectral function under external magnetic field and finite meson momentum}


\author{Min Zhou, Zhiyang Liu, Yvming Tian, Chonglong Xie, Guoyun Shao and Shijun Mao}
%\thanks{}
\email{maoshijun@mail.xjtu.edu.cn}
\affiliation{School of Physics, Xi'an Jiaotong University, Xi'an, Shaanxi 710049, P.R. China}


\date{\today}

\begin{abstract}
$q\bar{q}$ scattering phase shift in the $\pi^0$ channel $\Phi_{\pi^0}(\omega^2,\mathbf{k}_\perp^2,k^2_3)$ and ${\pi}^0$ meson spectral function $\rho_{\pi^0}(\omega^2,\mathbf{k}_\perp^2,k^2_3)$ under external magnetic field $eB$ and finite meson momentum $\mathbf{k}_\perp^2,k^2_3$ are studied in the framework of a two-flavor Nambu-Jona-Lasinio (NJL) model. The $q\bar{q}$ scattering phase shift in the $\pi^0$ channel $\Phi_{\pi^0}$ is closely related to $\pi^0$ spectral function $\rho_{\pi^0}$. We consider three situations, chiral broken phase ($T=\mu=0$), chiral restoration phase ($T>T_{pc},\ \mu=0$) and chiral restoration phase ($T=0,\ \mu>\mu_{pc}$). For $T=\mu=0$ and $T>T_{pc},\ \mu=0$ cases, ${\pi}^0$ meson spectral function $\rho_{\pi^0}$ shows a delta peak, several Breit-Wigner peaks and several non-Breit-Wigner peaks. The delta peak indicates the bound state of $\pi^0$ meson, and the Breit-Wigner peak means the resonant state of $\pi^0$ meson. For $T=0,\ \mu>\mu_{pc}$ case, Pauli blocking effect plays a role, which changes the inner structure of these Breit-Wigner peaks and non-Breit-Wigner peaks. Such multiple peak structure is caused by the external magnetic field. The $q\bar{q}$ scattering phase shift in the $\pi^0$ channel $\Phi_{\pi^0}$ shows a jump from $0$ to $\pi$ when $\pi^0$ meson is in bound state. When $\pi^0$ meson is in resonant state, $\Phi_{\pi^0}$ has the value $\pi/2$ and changes continuously. In large $\omega$ region, at the starting and end points of wide peaks of spectral function, $\Phi_{\pi^0}$ jumps abruptly (from $\pi$ to finite value or from finite value to $0$), and such jumps are caused by the external magnetic field. Finite momentum $\mathbf{k}_\perp^2$ or $k^2_3$ modifies the spectral function $\rho_{\pi^0}$ and scattering phase shift $\Phi_{\pi^0}$, which demonstrates the anisotropy in the system induced by external magnetic field.
\end{abstract}

%\keywords{}

\maketitle

\section{\label{sec:level1}Introduction}

The discovery of magnetars, the ultra-strong electromagnetic fields generated in non-central heavy-ion collisions, and the potential existence of primordial magnetic fields in the early universe have driven intense research on Quantum Chromodynamics (QCD) matter under extreme magnetic fields~\cite{Preis2013,Gatto2013,D'Elia2013,RevModPhys.88.025001,Cao2021,MIRANSKY20151,Shovkovy2013}, including the chiral restoration and deconfinement phase transitions, thermodynamic properties, and meson properties under external magnetic fields.

The study of hadron properties in a QCD medium is important for our understanding of strong interaction matter. For instance, the chiral symmetry breaking leads to the rich meson spectra, and the mass shift of hadrons will enhance or reduce their thermal production in relativistic heavy-ion collisions~\cite{Rapp2000,CASSING199965,PhysRevD.97.034026,Mao_2021,PhysRevD.96.034004,PhysRevD.86.025020,PhysRevD.109.016021,PhysRevD.103.076015,PhysRevD.107.074018,li2026massspectramotttransitions}. As the pseudo-Goldstone bosons of spontaneous chiral symmetry breaking, the properties of neutral pion $\pi^0$ at finite magnetic field, temperature, and density are widely investigated~\cite{Li:2022jqf,PhysRevD.86.085042,PhysRevD.99.056009,PhysRevD.99.056005,PhysRevD.103.094001,AVANCINI2017247,PhysRevD.104.094040,PhysRevD.108.054001,PhysRevD.83.025026}. In the previous studies, people focus on the static properties of $\pi^0$ meson with vanishing meson momentum. However, under external magnetic field, we should consider the anisotropy in the system.

In this paper, we systematically investigate the $q\bar{q}$ scattering phase shift in the $\pi^0$ channel and $\pi^0$ meson spectral function with external magnetic field and finite meson momentum within the framework of a two-flavor NJL model. We consider the conditions in chiral broken phase and in chiral symmetric phase. Anisotropy is studied by evaluating how these quantities depend individually on the transverse momentum $\mathbf{k}_\perp$ and longitudinal momentum $k_3$ of $\pi^0$ meson. In NJL model~\cite{PhysRev.122.345,RevModPhys.64.649,Volkov1993EffectiveCL,HATSUDA1994221,BUBALLA2005205}, quarks are elementary particles, and mesons are quantum fluctuations. The fluctuations' contribution to the thermodynamics of the quark-meson system can be expressed at the mean-field level for quarks and with the random phase approximation (RPA) for mesons, in terms of the bound states and scattering phase shifts of quark-antiquark pairs~\cite{HUFNER1994225,ZHUANG1994525,Xia_2019,Dubinin_2016,PhysRevD.96.094008,dubinin2013pionsigmamesondissociation}. This result appears to be more general than its application in the NJL model, as it resembles the Beth-Uhlenbeck formula for the second virial coefficient for a gas of non-relativistic particles and relativistic particles~\cite{UHLENBECK1936729,BETH1937915,dashen1969s,Schmidt:1990oyr}.

The rest of the paper is organized as follows. Section~\ref{sec:level2} establishes the theoretical framework. Section~\ref{sec:level3} presents the numerical results and discussions. Section~\ref{sec:level4} gives the summary and outlook. For a comprehensive study, we append the results with vanishing magnetic field in the Appendix.

\section{\label{sec:level2}Formalism}
The Lagrangian density of the two-flavor NJL model in an external magnetic field takes the following form \cite{PhysRev.122.345,Volkov1993EffectiveCL,RevModPhys.64.649,HATSUDA1994221,BUBALLA2005205},
\begin{equation}
	\label{lagragian}
	\mathcal{L}=\bar{\psi}(i\gamma_\nu D^\nu+{\hat \mu} \gamma_0-m_0) \psi+G[(\bar{\psi}\psi)^2+(\bar{\psi}i \gamma_5 \vec{\tau} \psi)^2].
\end{equation}
Here, $\psi$ represents the two-flavor quark field, and $m_0$ denotes the current quark mass, which explicitly breaks chiral symmetry. The covariant derivative $D^\nu=\partial^\nu+i Q A^\nu$ contains the coupling of charged quarks with $Q=\text{diag}(2e/3,-e/3)$ to a magnetic field, which is set along the $x_3$-axis $\mathbf{B}=(0,0,B)$, through the four-potential $A^\mu=(0,0,Bx_1,0)$ in the Landau gauge. ${\hat \mu}=\text{diag}(\mu_u,\mu_d)=\text{diag}(\mu,\mu)$ is the quark chemical potential matrix in flavor space. $G$ is the coupling constant in scalar and pseudoscalar channels.

\begin{figure}[htbp]
	\centering
	\includegraphics[width=0.45\textwidth]{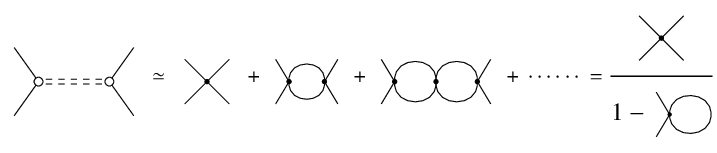}
	\caption{Effective interaction in random-phase approximation.}
	\label{feynmangraphic}
\end{figure}

In the NJL model, mesons are described as collective excitations. Beyond the mean-field approximation, the effective interaction between quarks is mediated by the mesons, and the meson propagator can be constructed with the random phase approximation (RPA), see Fig.~\ref{feynmangraphic}.

%%%%%%%%%%%%%%%%%%%%%%%%%%%%%%%%%%%%%%%%%%%%%%%%%%%%%%%%%%
%%%%%%%%%%%%%%%%%%%%%%%%%%%%%%%%%%%%%%%%%%%%%%%%%%%%%%%%%%%%%

With vanishing magnetic field, the approximate symmetry $SU_L(2)\otimes SU_R(2)$ is spontaneously broken to $SU_V(2)$ due to quark chiral condensate $\braket{\bar{\psi}\psi}$, and the pseudo-Goldstone bosons are $\pi^0$, $\pi^\pm$ mesons. The external magnetic field explicitly breaks $SU_V(2)$ symmetry down to $U_V(1)$, leaving $\pi^0$ meson as the only pseudo-Goldstone boson. Therefore, we focus on the properties of $\pi^0$ meson under external magnetic field in this work.

The $\pi^0$ meson propagator in momentum space with $k=(\omega,k_1, k_2, k_3)$ within RPA is given by
\begin{equation}
	\label{propa}
	D_{\pi^0}(k)=\frac{2G}{1-2G\Pi_{\pi^0}(k)}.
\end{equation}
The $\pi^0$ polarization function $\Pi_{\pi^0}(k)$, which is the quark bubble (see Fig.~\ref{feynmangraphic}), is defined in terms of the quark propagator matrix $S(x,y)=\text{diag}(S_u,S_d)$ as
\begin{equation}
\label{pldefinition}
	\begin{aligned}
		\Pi_{\pi^0}(k)&=-i\int d^4(x-x')e^{ik\cdot(x-x')}\\
		&\quad\times \mathrm{Tr}[i \tau_3\gamma_5 S(x,x') i \tau_3\gamma_5 S(x',x)],
	\end{aligned}
\end{equation}
with the trace taken over color, flavor and Dirac spaces.

The $\pi^0$ propagator determines the quark-antiquark scattering properties. Note that $\left(1-2G\Pi_{\pi^0}(k) \right)^{-1}$ is proportional to the ${\cal T}$-matrix for $q\bar{q}$ scattering in the $\pi^0$ channel~\cite{HUFNER1994225,ZHUANG1994525,Xia_2019}. The corresponding ${\cal S}$-matrix element can be expressed as
\begin{equation}
	\label{Smatrix1}
	{\cal S}_{\pi^0}(\omega^2,\mathbf{k}_\perp^2,k^2_3)=\frac{1-2 G \Pi_{\pi^0}((\omega-i\epsilon)^2,\mathbf{k}_\perp^2,k^2_3)}{1-2 G \Pi_{\pi^0}((\omega+i\epsilon)^2,\mathbf{k}_\perp^2,k^2_3)}.
\end{equation}
Being unimodular, it can be expressed in terms of the $q\bar{q}$ scattering phase shift $\Phi_{\pi^0}$ in the $\pi^0$ channel
\begin{equation}
	\label{Smatrix2}
	{\cal S}_{\pi^0}(\omega^2,\mathbf{k}_\perp^2,k^2_3)=e^{2 i \Phi_{\pi^0}(\omega^2,\mathbf{k}_\perp^2,k^2_3)},
\end{equation}
and
\begin{equation}
	\label{phasedefinition}
	\begin{aligned}
		&\Phi_{\pi^0}(\omega^2,\mathbf{k}_\perp^2,k^2_3)\\
		&=\arctan[\frac{2G\operatorname{Im}[\Pi_{\pi^0}((\omega+i\epsilon)^2,\mathbf{k}_\perp^2,k^2_3)]}{1-2G\operatorname{Re}[\Pi_{\pi^0}((\omega+i\epsilon)^2,\mathbf{k}_\perp^2,k^2_3)]}].
	\end{aligned}
\end{equation}
The $q{\bar q}$ scattering phase shift in $\pi^0$ channel $\Phi_{\pi^0}(\omega^2,\mathbf{k}_\perp^2,k^2_3)$ is closely related to the $\pi^0$ spectral function $\rho_{\pi^0}(\omega^2,\mathbf{k}_\perp^2,k^2_3)$,
\begin{equation}
	\rho_{\pi^0}(\omega^2,\mathbf{k}_\perp^2,k^2_3)
	=\frac{2G\sin[2\Phi_{\pi^0}(\omega^2,\mathbf{k}_\perp^2,k^2_3)]}{1-2G\operatorname{Re}[\Pi_{\pi^0}((\omega+i\epsilon)^2,\mathbf{k}_\perp^2,k^2_3)]},
\end{equation}
which is defined as
\begin{widetext}
	\begin{equation}
	\label{spec}
	\begin{aligned}
		&\rho_{\pi^0}(\omega^2,\mathbf{k}_\perp^2,k^2_3)=-2\operatorname{Im}[D_{\pi^0}((\omega+i\epsilon)^2,\mathbf{k}_\perp^2,k^2_3)]\\
		&=\frac{8G^2\operatorname{Im}[\Pi_{\pi^0}((\omega+i\epsilon)^2,\mathbf{k}_\perp^2,k^2_3)]}{(1-2G\operatorname{Re}[\Pi_{\pi^0}((\omega+i\epsilon)^2,\mathbf{k}_\perp^2,k^2_3)])^2+(2G\operatorname{Im}[\Pi_{\pi^0}((\omega+i\epsilon)^2,\mathbf{k}_\perp^2,k^2_3)])^2},
	\end{aligned}
\end{equation}
\end{widetext}
and encodes the properties of bound and resonant states.

Owing to the symmetry reduction from $O(1,3)$ to $O(3)$ by the medium ($T,\ \mu$), and subsequently to $O(2)$ by the external magnetic field, all quantities related to the $\pi^0$ meson depend separately on momentum $\omega^2$, $\mathbf{k}_\perp^2=k_1^2+k_2^2$ and $k^2_3$. Accordingly, they are written as functions of $(\omega^2,\mathbf{k}_\perp^2,k^2_3)$ throughout this manuscript.

Both the quark-antiquark scattering phase shift in $\pi^0$ channel and the $\pi^0$ meson spectral function are controlled by the polarization function. We now derive its analytical form, including the real and imaginary parts, and analyze the associated threshold conditions.

Based on the Leung--Ritus--Wang method \cite{RITUS1972555,LEUNG2006266,PhysRevC.79.035807,PhysRevC.82.065802,FUKUSHIMA2010311}, the quark propagator with flavor $f$ in coordinate space can be written as
\begin{align}
	S_f(x,y)&=i\sum_{n=0}^{\infty}\int \frac{d\tilde{p}}{(2\pi)^3}e^{-i\tilde{p}\cdot(x-y)}\nonumber\\
	&\quad \times P_n(x_1,p_2) D_f(\bar{p}) P_n(y_1,p_2),\\
	P_n(z,q)&=\frac{1}{2}[g_n^{s_f}(z,q)+I_ng_{n-1}^{s_f}(z,q)]\nonumber\\
	&\quad+\frac{is_f}{2}[g_n^{s_f}(z,q)-I_ng_{n-1}^{s_f}(z,q)]\gamma^1\gamma^2,\\
	D_f^{-1}(\bar{p})&=\gamma\cdot\bar{p}-m_q,
\end{align}
where $\tilde{p}=(p_0,0,p_2,p_3)$ is the Fourier transformed momentum, $\bar{p}=(p_0,0,-s_f\sqrt{2n|Q_fB|},p_3)$ is the conserved Ritus momentum with the sign factor $s_f=\text{sgn}(Q_fB)$ for $f=u,d$. $I_n=1-\delta_{n0}$ is Landau energy level factor. The magnetic field-dependent function $g_n^{s_f}(z,q)=\phi_n(z-s_fq/|Q_fB|)$ is expressed in terms of the Hermite polynomials $H_n(z)$ as $\phi_n(z)=\frac{|Q_fB|^{1/4}}
{\sqrt{2^n n!\sqrt{\pi}}}\times e^{-z^2|Q_fB|/2}H_n(z/|Q_fB|^{-1/2})$.

The quark mass $m_q=m_0-2G\braket{\bar{\psi}\psi}$ is determined by the gap equation at the mean-field level
\begin{equation}
	\label{gapeq}
	m_0-(1-2G{\cal I}_1)m_q=0,
\end{equation}
and
\begin{equation}
	\label{I1}
	\begin{aligned}
		{\cal I}_1&=N_c\sum_{f=u,d}\sum_{l=0}^{\infty}\int\frac{dq_3}{2\pi}  \alpha_l \frac{\left|Q_fB\right|}{2\pi} \\
		&\quad \times \frac{1-F(E_q+\mu)-F(E_q-\mu)}{E_q},
	\end{aligned}
\end{equation}
where $F(x)=1/\left(e^{x/T}+1\right)$ is Fermi-Dirac distribution function, $E_q=\sqrt{q_3^2+2l|Q_fB|+m_q^2}$ is quark energy in a magnetic field, the spin degeneracy is $\alpha_l=2-\delta_{l0}$, and the number of colors is $N_c=3$.

\begin{widetext}
Substituting the quark propagator $S_f(x,y)$ into Eq.\eqref{pldefinition}, and taking the analytic continuation $\omega\rightarrow\omega+i\epsilon$, the real part of the $\pi^0$ polarization function is
\begin{equation}
	\label{RepolareB}
	\begin{aligned}
		&{\Pi}^{{\cal{R}} e}(\omega^2,\mathbf{k}_\perp^2,k^2_3)=\operatorname{Re}[\Pi_{\pi^0}((\omega+i\epsilon)^2,\mathbf{k}_\perp^2,k^2_3)]=-N_c\sum_{f=u,d}\sum_{n,l=0}\mathrm{P.V.}\int \frac{dq_3}{2\pi}\frac{|Q_fB|}{2\pi}\\
		&\quad \left\{\frac{{\cal B}_{nlf}^+(\mathbf{k}_\perp^2)+{\cal B}_{nlf}^-(\mathbf{k}_\perp^2)}{4}\left(\frac{F(E_{q+k}-\mu)-F(-E_{q+k}-\mu)}{E_{q+k}}+\frac{F(E_q-\mu)-F(-E_q-\mu)}{E_q}\right)\right.\\
		&\quad  \quad +\frac{1}{4E_{q+k}E_q}\left(\frac{2(n+l)|Q_fB|-(\omega^2-k^2_3)}{2}\left[{\cal B}_{nlf}^+(\mathbf{k}_\perp^2)+{\cal B}_{nlf}^-(\mathbf{k}_\perp^2)\right]-2|Q_fB|\sqrt{nl}\left[{\cal B}_{nlf}^+(\mathbf{k}_\perp^2)-{\cal B}_{nlf}^-(\mathbf{k}_\perp^2)\right]\right)\\
		&\quad  \quad  \quad \times\left(\frac{F(E_q-\mu)-F(E_{q+k}-\mu)}{\omega+E_q-E_{q+k}}+ \frac{F(-E_q-\mu)-F(-E_{q+k}-\mu)}{\omega-E_q+E_{q+k}}\right.\\
		&\quad \quad  \quad  \quad \quad \left.\left.-\frac{F(E_q-\mu)-F(-E_{q+k}-\mu)}{\omega+E_q+E_{q+k}}-\frac{F(-E_q-\mu)-F(E_{q+k}-\mu)}{\omega-E_q-E_{q+k}}\right)\right\}.
	\end{aligned}
\end{equation}
$\mathrm{P.V.}$ denotes the Cauchy principal value integral when singularities exist in the cross terms $\propto\ \frac{1}{\omega\pm E_q\mp E_{q+k}}$ or non-cross term $\propto\ \frac{1}{\omega - E_q - E_{q+k}}$, with quark energy $E_q=\sqrt{2l|Q_fB|+q_3^2+m_q^2}$ and $E_{q+k}=\sqrt{2n|Q_fB|+(q_3+k_3)^2+m_q^2}$. The imaginary part of the $\pi^0$ polarization function is
\begin{equation}
	\label{ImpolareB}
	\begin{aligned}
		&\Pi^{{\cal {I}}m}(\omega^2,\mathbf{k}_\perp^2,k^2_3)=\operatorname{Im}[\Pi_{\pi^0}((\omega+i\epsilon)^2,\mathbf{k}_\perp^2,k^2_3)]=\pi N_c\sum_{f=u,d}\sum_{n,l=0}\int \frac{dq_3}{2\pi}\frac{|Q_fB|}{2\pi}\bigg\{\frac{1}{4E_{q+k}E_q}\bigg.\\
		&\times\bigg(\frac{2(n+l)|Q_fB|-(\omega^2-k^2_3)}{2}\left[{\cal B}_{nlf}^+(\mathbf{k}_\perp^2)+{\cal B}_{nlf}^-(\mathbf{k}_\perp^2)\right]-2|Q_fB|\sqrt{nl}\left[{\cal B}_{nlf}^+(\mathbf{k}_\perp^2)-{\cal B}_{nlf}^-(\mathbf{k}_\perp^2)\right]\bigg)\\
		&\times\bigg(\left[F(E_q-\mu)-F(E_{q+k}-\mu)\right]\delta(\omega+E_q-E_{q+k})+\left[F(-E_q-\mu)-F(-E_{q+k}-\mu)\right]\delta(\omega-E_q+E_{q+k})\bigg.\\
		&\quad \quad \bigg.\bigg.-\left[F(-E_q-\mu)-F(E_{q+k}-\mu)\right]\delta(\omega-E_q-E_{q+k})-\left[F(E_q-\mu)-F(-E_{q+k}-\mu)\right]\delta(\omega+E_q+E_{q+k}) \bigg)\bigg\}.
	\end{aligned}
\end{equation}
Here, the functions ${\cal B}_{nlf}^\pm(\mathbf{k}_\perp^2)$ are written as
\begin{equation}
	\label{functions}
	\begin{aligned}
		{\cal B}_{nlf}^\pm(\mathbf{k}_\perp^2)&=\exp\left(-\frac{\mathbf{k}_\perp^2}{2|Q_fB|}\right)\left(\frac{\mathbf{k}_\perp^2}{2|Q_fB|}\right)^{\Delta}\frac{{\zeta}!}{({\zeta}+\Delta)!}\left[L_{{\zeta}}^{(\Delta)}\left(\frac{\mathbf{k}_\perp^2}{2|Q_fB|}\right)\pm I_nI_l\sqrt{\frac{{\zeta}+\Delta}{{\zeta}}}L_{{\zeta}-1}^{(\Delta)}\left(\frac{\mathbf{k}_\perp^2}{2|Q_fB|}\right)\right]^2,
	\end{aligned}
\end{equation}
with $\zeta=\min(n,l)$, $\Delta=|n-l|$, and the generalized Laguerre polynomial $L_\zeta^{(\Delta)}(x)$ of convention $L_{-1}^{(\Delta)}(x)=0$ and $L_{\zeta}^{(0)}(0)=1$. ${\cal B}_{nlf}^\pm(\mathbf{k}_\perp^2)$ only depends on the momentum $\mathbf{k}_\perp^2$, and is not related to momentum $k^2_3$.

The singularities in Eq.\eqref{RepolareB} generate two classes of thresholds, namely the unitary thresholds~\cite{PhysRevD.109.016021,Le_Bellac_1996,skmg-ql8c} and Landau thresholds~\cite{PhysRevD.109.016021,Le_Bellac_1996,skmg-ql8c}. The unitary thresholds
\begin{equation}
	\label{thu}
	\omega^f_{U}(n,l)=\min(E_{q+k}+E_q)=\sqrt{2n|Q_fB|+\frac{k^2_3}{4}+m_q^2}+\sqrt{2l|Q_fB|+\frac{k^2_3}{4}+m_q^2},\ n,l=0,1,2,...
\end{equation}
arise from the non-cross term $\frac{1}{\omega-E_q-E_{q+k}}$ (see Table~\ref{thresholdunitary}). As $\omega\to \omega^f_{U}(n,l)^-$, the real part of polarization function $\Pi^{{\cal {R}}e}(\omega^2,\mathbf{k}_\perp^2,k^2_3)$  diverges to $+\infty$, and it remains finite for other values of $\omega$. The imaginary part of polarization function $\Pi^{{\cal {I}}m}(\omega^2,\mathbf{k}_\perp^2,k^2_3)$  diverges to $+\infty$ as $\omega\to \omega^f_{U}(n,l)^+$, and it remains finite for other values of $\omega$. Therefore, at unitary thresholds, the spectral function is zero. With increasing $\omega$, scattering phase shift $\Phi_{\pi^0}$ jumps from value $\pi$ to $\phi_U$ at unitary threshold. $\phi_U$ has a value greater (smaller) than $\pi/2$, when the sign of quantity $1-2G\Pi^{{\cal {R}}e}(\omega^2,\mathbf{k}_\perp^2,k^2_3)$ is negative (positive).

\begin{table}[htbp]
	\centering
	\renewcommand{\arraystretch}{1.3}
	\caption{Behavior near unitary thresholds with $eB \neq 0$ for any momentum.}
	\label{thresholdunitary}
	
	\begin{tabular}{ccccc}
		\hline
		&
		$\Pi^{{\cal {R}}e}(\omega^2,\mathbf{k}_\perp^2,k^2_3)$ &
		$\Pi^{{\cal {I}}m}(\omega^2,\mathbf{k}_\perp^2,k^2_3)$ &
		$\Phi_{\pi^0}(\omega^2,\mathbf{k}_\perp^2,k^2_3)$ &
		$\rho_{\pi^0}(\omega^2,\mathbf{k}_\perp^2,k^2_3)$
		\\
		\hline
		
		$\omega \to\omega^f_{U}(n,l)^-$ &
		$+\infty$ &
		finite &
		$\pi$ &
		0 \\[2mm]
		
		$\omega \to\omega^f_{U}(n,l)^+$ &
		finite &
		$+\infty$ &
		$\phi_U $ &
		$0$ \\
		
		\hline
	\end{tabular}
\end{table}

The Landau thresholds
\begin{equation}
	\label{thl}
	\omega^f_{L}(n,l)=\max|E_{q+k}-E_q|=\sqrt{k^2_3+(\sqrt{2n|Q_fB|+m_q^2}-\sqrt{2l|Q_fB|+m_q^2})^2},\ n,l=0,1,2,...
\end{equation}
arise from the cross terms $\frac{1}{\omega\pm E_q\mp E_{q+k}}$. Note that with $T=\mu=0$ or vanishing momentum $\mathbf{k}_\perp^2=k^2_3=0$, there are no cross terms and the Landau thresholds in Eq.\eqref{RepolareB}, due to the cancellation of the numerator in the cross terms $(\frac{F(E_q-\mu)-F(E_{q+k}-\mu)}{\omega+E_q-E_{q+k}},\  \frac{F(-E_q-\mu)-F(-E_{q+k}-\mu)}{\omega-E_q+E_{q+k}})$. When we meet the Landau thresholds, their contribution depends on the momentum. In the case $\mathbf{k}_\perp^2\neq 0, \ k^2_3=0$ (see Table~\ref{tab:thresholdkp}), as $\omega\to\omega^f_{L}(n,l)^+$, the real part $\Pi^{{\cal {R}}e}(\omega^2,\mathbf{k}_\perp^2,k^2_3)$ diverges to $-\infty$, and it remains finite for other values of $\omega$. The imaginary part $\Pi^{{\cal {I}}m}(\omega^2,\mathbf{k}_\perp^2,k^2_3)$ diverges to $+\infty$ as $\omega\to\omega^f_{L}(n,l)^-$, and it remains finite for other values of $\omega$. The spectral function is zero at these Landau thresholds. With increasing $\omega$, scattering phase shift $\Phi_{\pi^0}$ jumps from value $\phi_{L1}$ to $0$ at these Landau thresholds, and $\phi_{L1}$ has a value greater (smaller) than $\pi/2$ when the sign of quantity $1-2G\Pi^{{\cal {R}}e}(\omega^2,\mathbf{k}_\perp^2,k^2_3)$ is negative (positive). In the case $k^2_3\neq 0,\ \mathbf{k}_\perp^2=0$ (see Table~\ref{tab:thresholdk3}), the coefficients defined in Eq.\eqref{functions} become ${\cal B}_{nlf}^\pm(0)\propto\delta_{nl}$, and this leads to the Landau thresholds $\omega^f_{L}(n,n)$. There is no divergence in the polarization function at $\omega=\omega^f_{L}(n,n)$, and its real part $\Pi^{{\cal {R}}e}(\omega^2,\mathbf{k}_\perp^2,k^2_3)$ (imaginary part $\Pi^{{\cal {I}}m}(\omega^2,\mathbf{k}_\perp^2,k^2_3)$ ) has finite (zero) value. The spectral function is zero at these Landau thresholds. Scattering phase shift $\Phi_{\pi^0}$ is a continuous function, and has the value $\phi_{L2}=0$ $(\phi_{L2}=\pi)$ at $\omega=\omega^f_{L}(n,n)$ when the sign of quantity $1-2G\Pi^{{\cal {R}}e}(\omega^2,\mathbf{k}_\perp^2,k^2_3)$ is positive (negative).

\begin{table}[htbp]
	\centering
	\renewcommand{\arraystretch}{1.3}
	\caption{Behavior near Landau thresholds with $eB\neq 0$, $\mathbf{k}_\perp^2\neq0, \ k^2_3=0$.}
	\label{tab:thresholdkp}
	
	\begin{tabular}{ccccc}
		\hline
		&
		$\Pi^{{\cal {R}}e}(\omega^2,\mathbf{k}_\perp^2,k^2_3)$  &
		$\Pi^{{\cal {I}}m}(\omega^2,\mathbf{k}_\perp^2,k^2_3)$ &
		$\Phi_{\pi^0}(\omega^2,\mathbf{k}_\perp^2,k^2_3)$ &
		$\rho_{\pi^0}(\omega^2,\mathbf{k}_\perp^2,k^2_3)$
		\\
		\hline
		
		$\omega \to\omega^f_{L}(n,l)^-$ &
		finite &
		$+\infty$ &
		$\phi_{L1}$ &
		$0$ \\[2mm]
		
		$\omega \to\omega^f_{L}(n,l)^+$ &
		$-\infty$ &
		finite &
		0 &
		0 \\
		
		\hline
	\end{tabular}
\end{table}

\begin{table}[htbp]
	\centering
	\renewcommand{\arraystretch}{1.3}
	\caption{Behavior near Landau thresholds with $eB\neq 0$, $\mathbf{k}_\perp^2=0, \ k^2_3\neq0$.}
	\label{tab:thresholdk3}
	
	\begin{tabular}{ccccc}
		\hline
		&
		$\Pi^{{\cal {R}}e}(\omega^2,\mathbf{k}_\perp^2,k^2_3)$  &
		$\Pi^{{\cal {I}}m}(\omega^2,\mathbf{k}_\perp^2,k^2_3)$  &
		$\Phi_{\pi^0}(\omega^2,\mathbf{k}_\perp^2,k^2_3)$ &
		$\rho_{\pi^0}(\omega^2,\mathbf{k}_\perp^2,k^2_3)$
		\\
		\hline
		
		$\omega \to\omega^f_{L}(n,n)^-$ &
		finite &
		$0$ &
		$\phi_{L2}$&
		$0$ \\[2mm]
		
		$\omega \to\omega^f_{L}(n,n)^+$ &
		finite &
		0 &
		$\phi_{L2}$&
		0 \\
		
		\hline
	\end{tabular}
\end{table}

Since $|Q_u|=2|Q_d|$ for quarks, we use the flavor index $f=d$ when denoting unitary and Landau thresholds in the following context.

\end{widetext}

\section{\label{sec:level3}Numerical Results and discussion}

\begin{table}[htbp]
	\centering
	\renewcommand{\arraystretch}{1.3}
	\caption{The pseudo-critical temperature $T_{pc}$ (with vanishing quark chemical potential $\mu=0$) and pseudo-critical quark chemical potential $\mu_{pc}$ (with vanishing temperature $T=0$) for chiral restoration phase transition in the absence and presence of external magnetic field.}
	\label{tab:critical_values}
	
	\begin{tabular}{ccc}
		\hline
		$eB$            & $T_{pc}$ (GeV) & $\mu_{pc}$ (GeV)\\
		\hline
		$0$             & $0.157$        & $0.289$  \\
		$20m_{\pi}^2$   & $0.180$        & $0.235$  \\
		\hline
	\end{tabular}
\end{table}
Due to the non-renormalizability of the Nambu-Jona-Lasinio (NJL) model, regularization is necessary to handle ultraviolet divergences. We apply the Pauli-Villars regularization scheme~\cite{RevModPhys.64.649}. By fitting the quark chiral condensate $\braket{\bar{\psi}\psi}=-(0.25\ \text{GeV})^3$, the pion decay constant $f_{\pi}=93 \ \text{MeV}$, and the pion mass $m_{\pi}=134 \ \text{MeV}$ in vacuum, the parameters are determined to be $G = 3.44 \ \text{GeV}^{-2}$, $\Lambda = 1.127 \ \text{GeV}$, and $m_0 = 0.005 \ \text{GeV}$~\cite{PhysRevD.106.094017}.

With the parameters, the pseudo-critical temperature $T_{pc}$ (pseudo-critical quark chemical potential $\mu_{pc}$) with vanishing quark chemical potential $\mu=0$ (vanishing temperature $T=0$) is summarized in Table~\ref{tab:critical_values} with vanishing and finite magnetic field $eB=0,\ 20m_{\pi}^2$. In the following numerical calculations, we select representative value of temperature and quark chemical potential in both the chiral symmetry broken phase and chiral symmetry restored phase. For the chiral symmetry broken phase, we set $(eB,T,\mu)=(0,0,0)$ and $(20m_{\pi}^2,0,0)$. For the chiral symmetry restored phase, we use $(eB,T,\mu)=(0,0.2\text{GeV},0)$ and $(20m_{\pi}^2,0.2\text{GeV},0)$, as well as $(eB,T,\mu)=(0,0,0.35\text{GeV})$ and $(20m_{\pi}^2,0,0.25\text{GeV})$.

The results with finite magnetic field are discussed in this section, and the results with vanishing magnetic field are presented in the Appendix.

\subsection{Spectral function}
%%%%%%%%%%%%%%%%%%%%%%%%%%%%%%%%%%%%%%%%%%%%%%%%%%%%%%%%%%%%%%%%%%%%%%%%%%%%%
\begin{figure*}[htbp]
	\centering
	\includegraphics[width=0.45\textwidth]{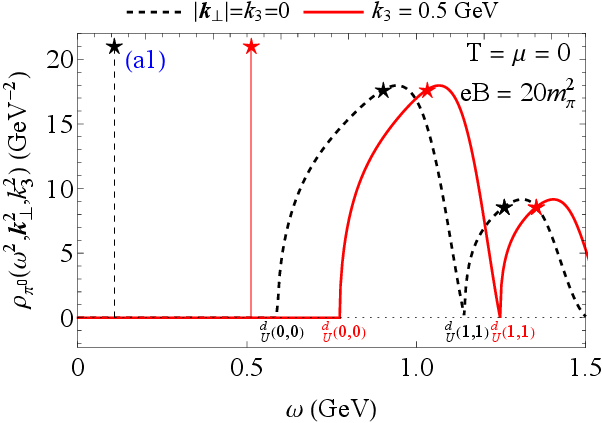}\includegraphics[width=0.45\textwidth]{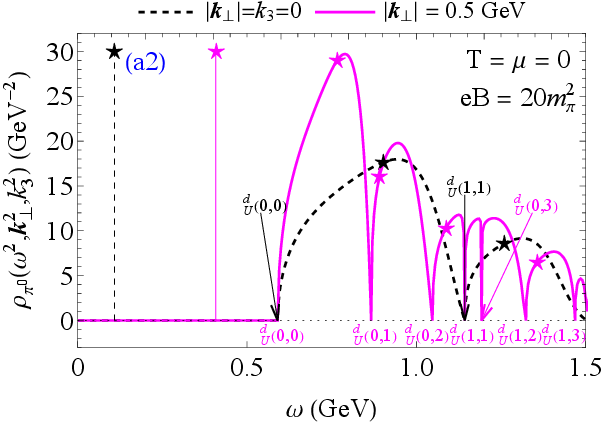}
	\vspace{2pt}
	\includegraphics[width=0.45\textwidth]{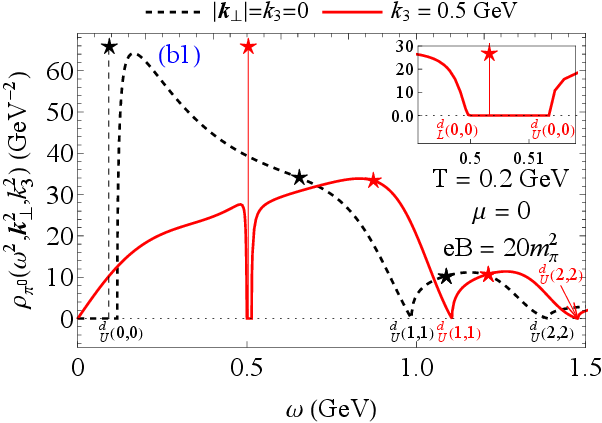}\includegraphics[width=0.45\textwidth]{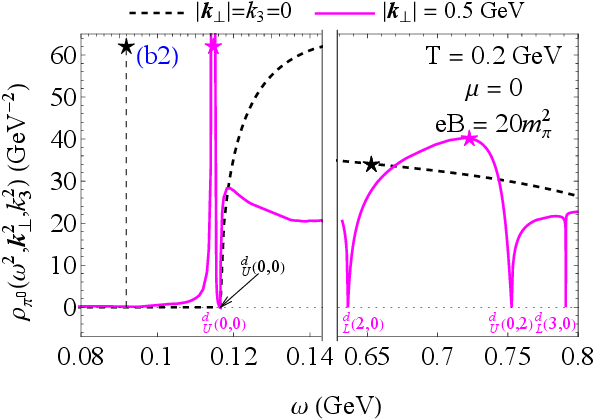}
	\vspace{2pt}
	\includegraphics[width=0.45\textwidth]{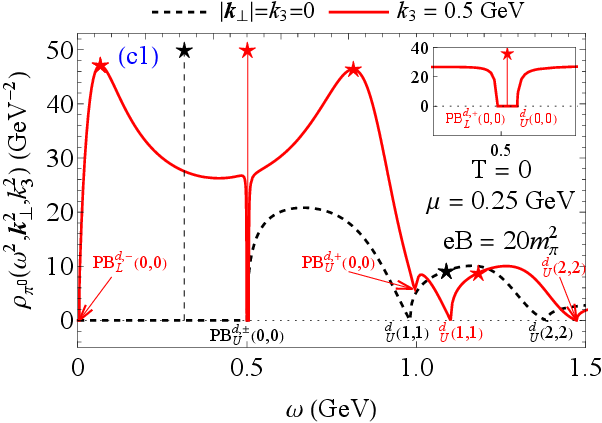}\includegraphics[width=0.45\textwidth]{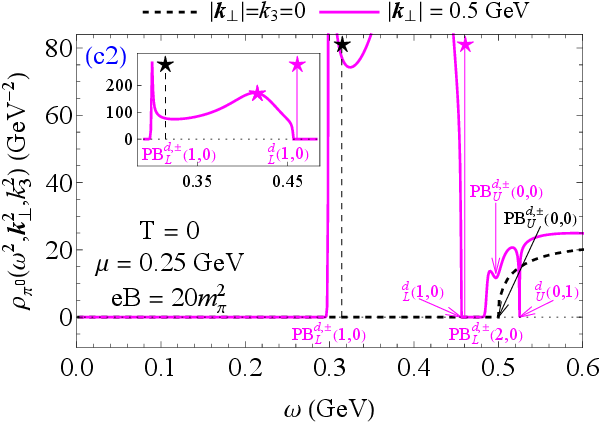}
	\caption{Spectral function of $\pi^0$ meson is shown as a function of $\omega$ for different $T$ and $\mu$ at fixed $eB=20m_{\pi}^2$. The black dashed, red solid, and magenta solid curves represent the case with zero momentum $|\mathbf{k}_\perp|=k_3=0$, with finite momentum $k_3=0.5$ GeV, $|\mathbf{k}_\perp|=0$, and with finite momentum $|\mathbf{k}_\perp|=0.5$ GeV, $k_3=0$, respectively. Here $_{U/L}^d(n,l)$ denotes unitary and Landau thresholds, and ${\text{PB}}_{U/L}^{d,\pm}(n,l)$ denotes Pauli-blocking thresholds. The star marks the location of the solutions of pole equation $1 - 2G\Pi^{{\cal {R}}e}(\omega^2,\mathbf{k}_\perp^2,k^2_3)=0$.}
	\label{spectral}
\end{figure*}
%%%%%%%%%%%%%%%%%%%%%%%%%%%%%%%%%%%%%%%%%%%%%%%%%%%%%%%%%%%%%%%%%%%%%%%%%%%%%

Figure~\ref{spectral} plots the spectral function of $\pi^0$ meson with finite magnetic field $(eB=20m_{\pi}^2)$. We focus on the effect of finite momentum ($k^2_3 \neq 0$ or $\mathbf{k}_\perp^2 \neq 0$) in chiral symmetry broken phase ($T=\mu=0$) and in chiral symmetry restored phase ($T>T_{pc}$ or $\mu >\mu_{pc}$).

\subsubsection{$T=\mu=0$}

For $T=\mu=0$ and $k^2_3 =\mathbf{k}_\perp^2= 0$ (see black dashed line in Fig.~\ref{spectral}(a1) and Fig.~\ref{spectral}(a2)), the $\pi^0$ spectral function shows a delta peak and several peaks of Breit-Wigner type. The delta peak indicates the bound state of $\pi^0$ meson, and the Breit-Wigner peak means the resonant state of $\pi^0$ meson. The location of the delta peak can be identified as the pole of $\pi^0$ propagator in Eq.\eqref{propa} $1-2G\Pi^{{\cal {R}}e}(\omega^2,0, 0)=0$, which is marked by the star. The imaginary part $\Pi^{{\cal {I}}m}(\omega^2,0, 0)$ is zero for $\omega<\omega^d_{U}(0,0)$. Breit-Wigner peaks appear between $\omega^d_{U}(n,n)$ and $\omega^d_{U}(n+1,n+1)$ for $n=0,1,2,...$. With $T=\mu=0$, the cross terms and the corresponding Landau thresholds do not make any contribution in Eq.\eqref{RepolareB} and Eq.\eqref{ImpolareB}, and we only need to consider the non-cross terms and the corresponding unitary thresholds. Furthermore, with $\mathbf{k}_\perp^2 = 0$, the coefficients defined in Eq.\eqref{functions} become ${\cal B}_{nlf}^\pm(0)\propto\delta_{nl}$, and hence the unitary thresholds with $n=l$ determine the appearance of Breit-Wigner peaks. There exists deviation between the locations of the star and the summit of Breit-Wigner peak, due to the nonzero imaginary part $\Pi^{{\cal {I}}m}(\omega^2,0, 0)\neq 0$ with $\omega>\omega^d_{U}(0,0)$.

The red line in Fig.~\ref{spectral}(a1) depicts the $\pi^0$ spectral function with finite momentum $k^2_3 \neq0,\ \mathbf{k}_\perp^2= 0$ for $T=\mu=0$. With magnetic field along $x_3$ direction, the boost invariance along $k_3$ direction is maintained. The $\pi^0$ spectral function with finite momentum $k^2_3 \neq0,\ \mathbf{k}_\perp^2= 0$ (red line) can be obtained by a shift $\omega \rightarrow \sqrt{\omega^2+k^2_3}$ in $\pi^0$ spectral function with $k^2_3 =\mathbf{k}_\perp^2= 0$ (black line).

The magenta line in Fig.~\ref{spectral}(a2) plots the $\pi^0$ spectral function with finite momentum $ \mathbf{k}_\perp^2 \neq 0,\ k^2_3 =0$ for $T=\mu=0$, which shows a delta peak and several peaks of Breit-Wigner type and non-Breit-Wigner type. With $\mathbf{k}_\perp^2 \neq 0$, the unitary thresholds with $n\neq l$ also lead to peaks. It is noticeable that not all the peaks correspond to the solutions of pole equation $1-2G\Pi^{{\cal {R}}e}(\omega^2,\mathbf{k}_\perp^2, 0)=0$, marked by the stars. In the context, the wide peak associated with the star mark is named as Breit-Wigner peak, and otherwise it is called non-Breit-Wigner peak. The $\pi^0$ spectral function with finite momentum $ \mathbf{k}_\perp^2 \neq 0,\ k^2_3 =0$ (magenta line) can not be obtained by a shift $\omega \rightarrow \sqrt{\omega^2+\mathbf{k}_\perp^2}$ in $\pi^0$ spectral function with $k^2_3 =\mathbf{k}_\perp^2= 0$ (black line), which represents the anisotropy induced by the external magnetic field.

\subsubsection{$T\neq 0,\ \mu=0$}

Figure~\ref{spectral}(b1) and Figure~\ref{spectral}(b2) show the $\pi^0$ spectral function in chiral restoration phase with $T=0.2{\text{GeV}},\ \mu=0$. With vanishing momentum $k^2_3 =\mathbf{k}_\perp^2= 0$ (see black lines), the $\pi^0$ spectral function shows a delta peak and several peaks of Breit-Wigner type. Apparent deviation between the location of the star and the summit of Breit-Wigner peak happens with $\omega^d_{U}(0,0)<\omega<\omega^d_{U}(1,1)$. With vanishing momentum $k^2_3 =\mathbf{k}_\perp^2= 0$, the cross terms and the Landau thresholds do not contribute to the polarization function, and only the non-cross terms and the unitary thresholds with $n=l$ make contributions. Further increasing $\omega$, multiple peaks appear with $\omega^d_{U}(n,n)<\omega<\omega^d_{U}(n+1,n+1),\ n=1,2,3...$. 

With finite momentum $k^2_3 \neq0,\ \mathbf{k}_\perp^2= 0$ (red line in Fig.~\ref{spectral}(b1)), the cross terms and the Landau thresholds also make contributions to the polarization function. Since the condition $\mathbf{k}_\perp^2= 0$ requires $n=l$ for quark Landau levels in Eq.\eqref{RepolareB} and Eq.\eqref{ImpolareB}, we meet only one Landau threshold $\omega^d_{L}(n,n)=\omega^d_{L}(0,0)=\left|k_3\right|, n=0,1,2,...$ and many unitary thresholds $\omega^d_{U}(n,n),\ n=0,1,2,...$. We observe a non-Breit-Wigner peak with $0<\omega<\omega^d_{L}(0,0)$ in the $\pi^0$ spectral function. For larger $\omega$, it shows a delta peak with $\omega^d_{L}(0,0)<\omega<\omega^d_{U}(0,0)$, and several peaks of Breit-Wigner type with $\omega^d_{U}(n,n)<\omega<\omega^d_{U}(n+1,n+1),\ n=0,1,2,...$.

However, with $ \mathbf{k}_\perp^2 \neq 0,\ k^2_3 =0$, many Landau thresholds $\omega^d_{L}(n,l), n,l=0,1,2,...$ and many unitary thresholds $\omega^d_{U}(n,l),\ n,l=0,1,2,...$ should be considered in the polarization function Eq.\eqref{RepolareB} and Eq.\eqref{ImpolareB}. This will lead to many more peaks in the $\pi^0$ spectral function. In Fig.~\ref{spectral}(b2), two segments of the results are plotted to illustrate the property of $\pi^0$ spectral function with finite temperature $(T\neq 0,\ \mu=0)$ and finite momentum $ (\mathbf{k}_\perp^2 \neq 0,\ k^2_3 =0)$ (see magenta lines). In low $\omega$ region, a sharp Breit-Wigner peak shows up with $\omega<\omega^d_{U}(0,0)$, which is accompanied by a lower non-Breit-Wigner peak nearby. With larger $\omega$, for instance $0.6{\text {GeV}}<\omega<0.8{\text {GeV}}$, several wide peaks appear between the unitary threshold and the Landau threshold. As shown in Table~\ref{thresholdunitary},~\ref{tab:thresholdkp},~\ref{tab:thresholdk3}, spectral function is always zero at unitary and Landau thresholds, and thus peak structure will appear between them, which rely on the numerical calculations.

\subsubsection{$T=0,\ \mu \neq 0$}
\begin{widetext}
With zero temperature and finite quark chemical potential, the Fermi-Dirac distribution in Eq.\eqref{RepolareB} and in Eq.\eqref{ImpolareB} reduces to a step function, $F(E_q-\mu)\to \Theta(\mu-E_q)$. This introduces additional Pauli blocking effect~\cite{BUBALLA2005205,HUFNER1994225,ZHUANG1994525}. We define the Pauli-blocking (PB) thresholds
\begin{equation}
	\text{PB}^{f,\pm}_{U}(n,l)=\sqrt{\mu^2+k^2_3\pm2k_3\sqrt{\mu^2-m_q^2-2(\zeta+\Delta)|Q_fB|}-2\Delta|Q_fB|}+\mu,\ n,l=0,1,2,...
\label{PBU}
\end{equation}
arising from the non-cross term, and
\begin{equation}
	\text{PB}^{f,\pm}_{L}(n,l)=\sqrt{\mu^2+k^2_3\pm2k_3\sqrt{\mu^2-m_q^2-2\zeta|Q_fB|}+2\Delta|Q_fB|}-\mu,\ n,l=0,1,2,...
\label{PBL}
\end{equation}
arising from the cross terms, which form another threshold condition for the emergence of imaginary part of polarization function in Eq.\eqref{ImpolareB}. It should be mentioned that PB thresholds do not induce any divergence in the real and imaginary parts of polarization function, and only PB thresholds with positive value play the role in the calculation. Since $|Q_u|=2|Q_d|$ for quarks, we use the flavor index $f=d$ when denoting PB thresholds in the following context.

\end{widetext}

Figure~\ref{spectral}(c1) and Figure~\ref{spectral}(c2) show the $\pi^0$ spectral function in chiral restoration phase with $T=0,\ \mu=0.25{\text{GeV}}$. With vanishing momentum $k^2_3 =\mathbf{k}_\perp^2= 0$ (see black lines), the $\pi^0$ spectral function shows a delta peak and several peaks of Breit-Wigner type and non-Breit-Wigner type. Due to the Pauli blocking effect, the wide peak starts to appear at $\omega=\text{PB}^{d,\pm}_{U}(n,n)=2\mu$, instead of at unitary threshold $\omega=\omega^d_{U}(0,0)$, which is the case in Figure~\ref{spectral} (a1,a2) and (b1,b2) black lines. Note that with $\mathbf{k}_\perp^2= k^2_3 =0$, we only have one PB threshold $\text{PB}^{f,\pm}_{U}(n,n)=2\mu, \ n=0,1,2,...$ Other wide peaks appear with $\omega^d_{U}(n,n)<\omega<\omega^d_{U}(n+1,n+1)$ for $n=1,2,3,...$.

Figure~\ref{spectral}(c1) red line plots the $\pi^0$ spectral function with finite momentum $k^2_3 \neq 0,\ \mathbf{k}_\perp^2= 0$. The PB thresholds become $\text{PB}^{d,\pm}_{U}(n,n)=\sqrt{\mu^2+k^2_3\pm2k_3\sqrt{\mu^2-m_q^2-2n|Q_fB|}}+\mu$ and $\text{PB}^{d,\pm}_{L}(n,n)=\sqrt{\mu^2+k^2_3\pm2k_3\sqrt{\mu^2-m_q^2-2n|Q_fB|}}-\mu$. With the considered numerical condition, $eB=20m_{\pi}^2,\ \mu=0.25{\text{GeV}}$, four positive PB thresholds exist, with $\text{PB}^{d,-}_{L}(0,0)<\text{PB}^{d,+}_{L}(0,0)<\text{PB}^{d,-}_{U}(0,0)<\text{PB}^{d,+}_{U}(0,0)$. In the spectral function, we observe a Breit-Wigner peak with $\text{PB}^{d,-}_{L}(0,0)<\omega<\text{PB}^{d,+}_{L}(0,0)$ and a delta peak with $\text{PB}^{d,+}_{L}(0,0)<\omega<\omega^{d}_{U}(0,0)$ ($\text{PB}^{d,-}_{U}(0,0)<\omega^{d}_{U}(0,0)$). The next Breit-Wigner peak appears with $\omega^{d}_{U}(0,0)<\omega<\omega^{d}_{U}(1,1)$, which contains a dip structure at $\omega=\text{PB}^{d,+}_{U}(0,0)$. For larger $\omega$, wide peaks appear with $\omega^{d}_{U}(n,n)<\omega<\omega^{d}_{U}(n+1,n+1)$ for $n=1,2,3,...$.

Figure~\ref{spectral}(c2) magenta line plots the $\pi^0$ spectral function with finite momentum $ \mathbf{k}_\perp^2 \neq 0,\ k^2_3 = 0$. In this case, we have $\text{PB}^{d,+}_{U}(n,l)=\text{PB}^{d,-}_{U}(n,l)=\sqrt{\mu^2-2|n-l||Q_dB|}+\mu$ and $\text{PB}^{d,+}_{L}(n,l)=\text{PB}^{d,-}_{L}(n,l)=\sqrt{\mu^2+2|n-l||Q_dB|}-\mu$. With the considered numerical condition, $eB=20m_{\pi}^2,\ \mu=0.25{\text{GeV}}$, we have only one positive $\text{PB}^{d,\pm}_{U}(n,n)=2\mu,\ n=0,1,2,...$, and many $\text{PB}^{d,\pm}_{L}(n,l),\ n,l=0,1,2,...$. With $\text{PB}^{d,\pm}_{L}(1,0)<\omega<\omega^{d}_{L}(1,0)$, a non-Breit-Wigner peak and a Breit-Wigner peak appear in $\pi^0$ spectral function. After that, we observe a delta peak. With $\text{PB}^{d,\pm}_{L}(2,0)<\omega<\omega^{d}_{U}(0,1)$, there is a non-Breit-Wigner peak associated with a dip at $\omega=\text{PB}^{d,\pm}_{U}(0,0)=2\mu$. With larger $\omega$, we have wide peaks between unitary and Landau thresholds in the $\pi^0$ spectral function, and the PB thresholds may modify the inner structure of these peaks, which rely on the numerical calculations. Since we meet too many unitary, Landau and PB thresholds, which make difficult to show the spectral function in the whole $\omega$ region clearly. We just plot the $\pi^0$ spectral function with $0<\omega<0.6{\text{GeV}}$ to characterize its property.

\subsection{Scattering phase shift}

%%%%%%%%%%%%%%%%%%%%%%%%%%%%%%%%%%%%%%%%%%%%%%%%%%%%%%%%%%%%%%%%%%%%%%%%%%%%%
\begin{figure*}[htbp]
		\centering
	\includegraphics[width=0.5\textwidth]{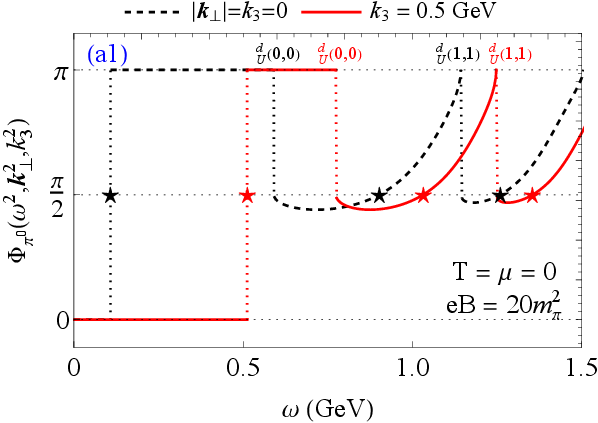}\includegraphics[width=0.5\textwidth]{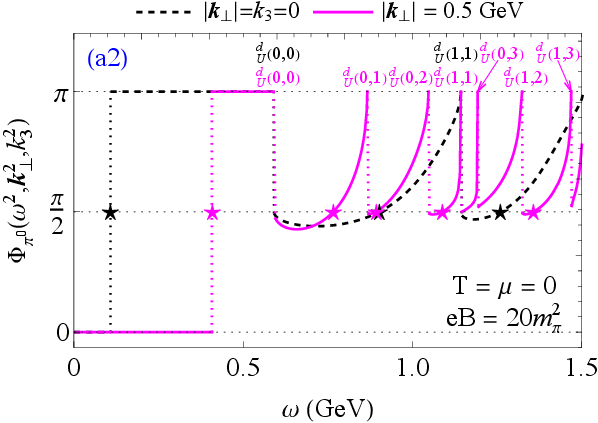}
	\vspace{2pt}
	\includegraphics[width=0.5\textwidth]{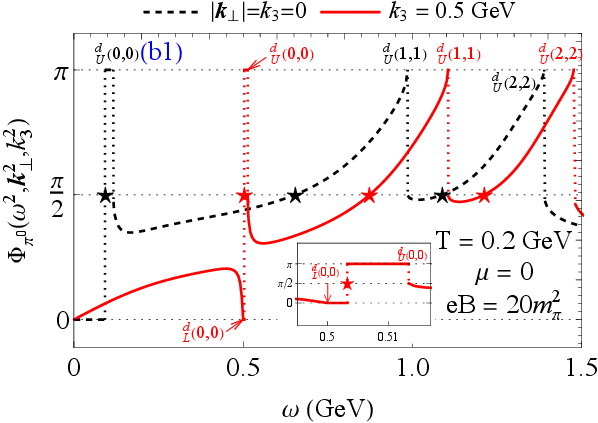}\includegraphics[width=0.5\textwidth]{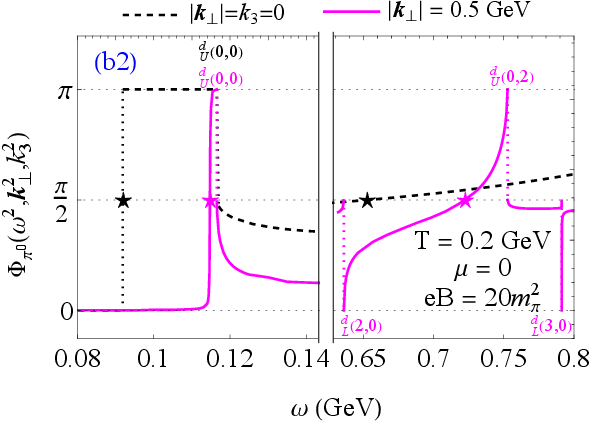}
	\vspace{2pt}
	\includegraphics[width=0.5\textwidth]{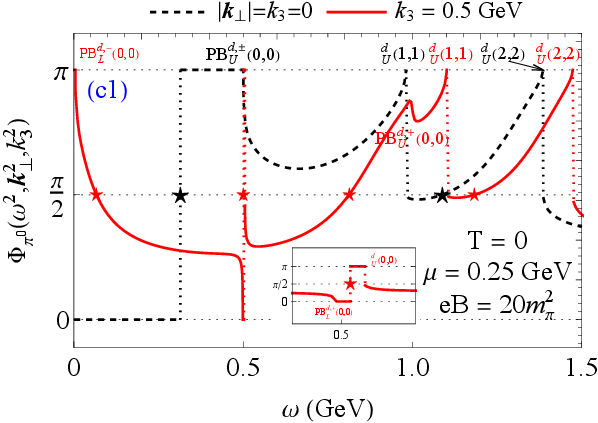}\includegraphics[width=0.5\textwidth]{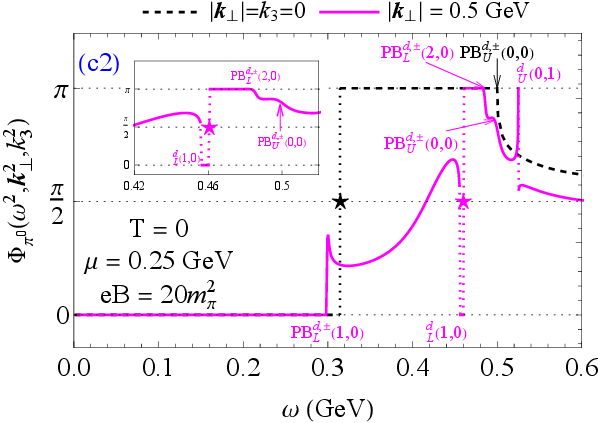}
	\caption{The $q\bar{q}$ scattering phase shift in the $\pi^0$ channel $\Phi_{\pi^0}$ is shown as a function of $\omega$ for different $T$ and $\mu$ at fixed $eB=20m_{\pi}^2$. The black dashed, red solid, and magenta solid curves represent the case with zero momentum $|\mathbf{k}_\perp|=k_3=0$, with finite momentum $k_3=0.5$ GeV, $|\mathbf{k}_\perp|=0$, and with finite momentum $|\mathbf{k}_\perp|=0.5$ GeV, $k_3=0$, respectively. Here $_{U/L}^d(n,l)$ denotes unitary and Landau thresholds, and ${\text{PB}}_{U/L}^{d,\pm}(n,l)$ denotes Pauli-blocking thresholds. The star marks the location of the solutions $\Phi_{\pi^0}(\omega^2,\mathbf{k}_\perp^2,k^2_3)=\pi/2$.}
	\label{phase}
\end{figure*}
%%%%%%%%%%%%%%%%%%%%%%%%%%%%%%%%%%%%%%%%%%%%%%%%%%%%%%%%%%%%%%%%%%%%%%%%%%%%%

Figure~\ref{phase} plots the $q\bar{q}$ scattering phase shift in $\pi^0$ channel $\Phi_{\pi^0}$ with finite magnetic field $(eB=20m_{\pi}^2)$. We focus on the effect of finite momentum ($k^2_3 \neq 0$ or $\mathbf{k}_\perp^2 \neq 0$) in chiral symmetry broken phase ($T=\mu=0$) and in chiral symmetry restored phase ($T>T_{pc}$ or $\mu >\mu_{pc}$).

According to the definition of $q\bar{q}$ scattering phase shift in Eq.\eqref{phasedefinition}, when we meet the pole of $\pi^0$ propagator $1 - 2G\Pi^{{\cal {R}}e}(\omega^2_{\delta},\mathbf{k}_\perp^2,k^2_3)=0$ with vanishing imaginary part of polarization function $\Pi^{{\cal {I}}m} (\omega^2_{\delta},\mathbf{k}_\perp^2,k^2_3)=0$, we obtain a step function $\Theta(x)$
\begin{equation}
\label{phasedefinition1}
		\Phi_{\pi^0}(\omega^2,\mathbf{k}_\perp^2,k^2_3)=\pi\Theta(\omega^2-\omega_{\delta}^2),
\end{equation}
with $\omega$ near $\omega_{\delta}$.
At the pole of $\pi^0$ propagator $1 - 2G\Pi^{{\cal {R}}e}(\omega^2_{\text{BW}},\mathbf{k}_\perp^2,k^2_3)=0$ with nonvanishing imaginary part of polarization function $\Pi^{{\cal {I}}m} (\omega^2_{\text{BW}},\mathbf{k}_\perp^2,k^2_3)\neq 0$, we have
\begin{equation}
\label{phasedefinition2}
		\Phi_{\pi^0}(\omega^2,\mathbf{k}_\perp^2,k^2_3)=\pi/2,
\end{equation}
with $\omega=\omega_{\text{BW}}$.
Therefore, we mark a star at $\Phi_{\pi^0}=\pi/2$ in all panels of Figure~\ref{phase}.

At unitary and Landau thresholds, jumps of scattering phase shift happen, as discussed in the previous section, see Table~\ref{thresholdunitary},~\ref{tab:thresholdkp},~\ref{tab:thresholdk3}.

\subsubsection{$T=\mu=0$}
Figure~\ref{phase}(a1) and Figure~\ref{phase}(a2) black lines depict the $q\bar{q}$ scattering phase shift in $\pi^0$ channel with $T=\mu=0$ and $k^2_3=\mathbf{k}_\perp^2= 0$. With increasing $\omega$, the scattering phase shift $\Phi_{\pi^0}$ shows a jump from $0$ to $\pi$ at $\omega=\omega_{\delta}$. With $\omega_{\delta}+0^+<\omega<\omega^d_U(0,0)-0^+$, we have $\Phi_{\pi^0}=\pi$. A sudden jump from $\pi$ to $\phi_U<\pi/2$ is observed at unitary threshold $\omega=\omega^d_U(0,0)$. After that, $\Phi_{\pi^0}$ decreases and then increases, which crosses $\Phi_{\pi^0}=\pi/2$ at $\omega=\omega_{\text{BW}}$ and reaches $\Phi_{\pi^0}=\pi$ at another unitary threshold $\omega=\omega^d_U(1,1)-0^+$. Similar behavior is observed with $\omega^d_U(n,n)<\omega<\omega^d_U(n+1,n+1),\ n=0,1,2,...$. 

Figure~\ref{phase}(a1) red lines plot the $q\bar{q}$ scattering phase shift in $\pi^0$ channel with $T=\mu=0$ and $k^2_3\neq 0, \ \mathbf{k}_\perp^2= 0$. It recovers the results in black line with the replacement $\omega \rightarrow \sqrt{\omega^2-k^2_3}$.

Figure~\ref{phase}(a2) magenta lines plot the $q\bar{q}$ scattering phase shift in $\pi^0$ channel with $T=\mu=0$ and $k^2_3= 0, \ \mathbf{k}_\perp^2\neq 0$. With increasing $\omega$, the scattering phase shift $\Phi_{\pi^0}$ shows a jump from $0$ to $\pi$ at $\omega=\omega_{\delta}$. With $\omega_{\delta}+0^+<\omega<\omega^d_U(0,0)-0^+$, we have $\Phi_{\pi^0}=\pi$. A sudden jump from $\pi$ to $\phi_U<\pi/2$ is observed at unitary threshold $\omega=\omega^d_U(0,0)+0^+$. After that, $\Phi_{\pi^0}$ decreases and then increases, which crosses $\Phi_{\pi^0}=\pi/2$ at $\omega=\omega_{\text{BW}}$ and reaches $\Phi_{\pi^0}=\pi$ at unitary threshold $\omega=\omega^d_U(0,1)-0^+$. Note that, with $k^2_3= 0, \ \mathbf{k}_\perp^2\neq 0$, we have unitary thresholds $\omega^d_U(n,l),\ n,l=0,1,2,...$. Different behavior is observed for $\omega$ in the region between different unitary thresholds. The value of $\phi_U$ at $\omega=\omega^d_U(n,l)+0^+$ can be smaller (larger) than $\pi/2$, when the sign of quantity $1-2G\Pi^{{\cal {R}}e}(\omega^2,\mathbf{k}_\perp^2,k^2_3)$ is negative (positive). When $\phi_U>\pi/2$, for instance at $\omega=\omega^d_U(1,1)+0^+$ and $\omega=\omega^d_U(0,3)+0^+$, $\Phi_{\pi^0}$ monotonically increases up to $\pi$ until the next unitary threshold. $\Phi_{\pi^0}$ in magenta lines can not recover the results in black lines with the replacement $\omega \rightarrow \sqrt{\omega^2-\mathbf{k}_\perp^2}$.

\subsubsection{$T\neq 0,\ \mu=0$}
Figure~\ref{phase}(b1) and Figure~\ref{phase}(b2) black lines depict the $q\bar{q}$ scattering phase shift in $\pi^0$ channel with $T\neq 0,\ \mu=0$ and $k^2_3=\mathbf{k}_\perp^2= 0$, which demonstrate similar behavior as in Figure~\ref{phase}(a1) and Figure~\ref{phase}(a2) black lines. Note that the values of $\omega_{\delta},\ \omega_{\text{BW}}$ and $\omega^d_U(n,n)$ are varied by temperature.

Figure~\ref{phase}(b1) red lines plot the $q\bar{q}$ scattering phase shift in $\pi^0$ channel with $T\neq 0,\ \mu=0$ and $k^2_3\neq 0, \ \mathbf{k}_\perp^2= 0$. We meet only one Landau threshold $\omega^d_L(0,0)$, which is smaller than $\omega_{\delta}$. With $0<\omega<\omega^d_L(0,0)$, scattering phase shift $\Phi_{\pi^0}$ firstly increases and then decreases to zero at $\omega=\omega^d_L(0,0)$, which changes continuously. After that, the red lines indicate similar properties as the black line. Apparently, it can not recover the results in black line with the replacement $\omega \rightarrow \sqrt{\omega^2-{k}^2_3}$, due to the breaking of Lorentz symmetry by temperature.

Figure~\ref{phase}(b2) magenta lines plot the $q\bar{q}$ scattering phase shift in $\pi^0$ channel with $T\neq 0,\ \mu=0$ and $k^2_3= 0, \ \mathbf{k}_\perp^2\neq 0$. Since we meet many unitary and Landau thresholds, we plot two segments of the results, as in Figure~\ref{spectral}(b2). With increasing $\omega$ ($0<\omega<0.14{\text{GeV}}$), $\Phi_{\pi^0}$ increases continuously from zero to $\pi$ at $\omega=\omega^d_U(0,0)-0^+$, which crosses $\pi/2$ at $\omega=\omega_{\text{BW}}<\omega^d_U(0,0)$. With $\omega=\omega^d_U(0,0)+0^+$, $\Phi_{\pi^0}$ jumps from $\pi$ to $\phi_U<\pi/2$. In the region $0.6{\text{GeV}}<\omega<0.8{\text{GeV}}$, $\Phi_{\pi^0}$ jumps from $\phi_L<\pi/2$ to $0$ at $\omega=\omega^d_L(2,0)-0^{+}$. After that, it increases from $0$ to $\pi$ continuously with $\omega^d_L(2,0)+0^{+}<\omega<\omega^d_U(0,2)-0^{+}$, which passes through $\pi/2$ at $\omega=\omega_{\text{BW}}$. At $\omega=\omega^d_U(0,2)+0^{+}$, $\Phi_{\pi^0}$ jumps from $\pi$ to $\phi_U<\pi/2$. With $\omega^d_U(0,2)+0^{+}<\omega<\omega^d_L(3,0)-0^{+}$, $\Phi_{\pi^0}$ slightly decreases and then slightly increases, and finally jumps to $0$ at $\omega=\omega^d_L(3,0)+0^{+}$.

\subsubsection{$T= 0,\ \mu \neq 0$}
With zero temperature and finite quark chemical potential, in addition to the unitary and Landau thresholds, we meet PB thresholds, due to the Pauli-blocking effect. It should be mentioned that at PB thresholds, no jumps of scattering phase shift happen.

Figure~\ref{phase}(c1) and Figure~\ref{phase}(c2) black lines depict the $q\bar{q}$ scattering phase shift in $\pi^0$ channel with $T=0,\ \mu \neq 0$ and $k^2_3=\mathbf{k}_\perp^2= 0$. With increasing $\omega$, the scattering phase shift $\Phi_{\pi^0}$ shows a jump from $0$ to $\pi$ at $\omega=\omega_{\delta}$. With $\omega_{\delta}+0^+<\omega \leq {\text {PB}}^{d,\pm}_U(0,0)$ ($\omega^d_U(0,0)<{\text {PB}}^{d,\pm}_U(0,0)$), we have $\Phi_{\pi^0}=\pi$. With ${\text {PB}}^{d,\pm}_U(0,0)\leq \omega<\omega^d_U(1,1)-0^+$, $\Phi_{\pi^0}$ first decreases from $\pi$ and then increases up to $\pi$. A sudden jump from $\pi$ to $\phi_U<\pi/2$ is observed at $\omega=\omega^d_U(1,1)+0^+$. After that, $\Phi_{\pi^0}$ slightly decreases and then increases, which crosses $\Phi_{\pi^0}=\pi/2$ at $\omega=\omega_{\text{BW}}$ and reaches $\Phi_{\pi^0}=\pi$ at $\omega=\omega^d_U(2,2)-0^+$. Similar behavior is observed in $\omega^d_U(n,n)<\omega<\omega^d_U(n+1,n+1),\ n=1,2,3,...$.

Figure~\ref{phase}(c1) red lines plot the $q\bar{q}$ scattering phase shift in $\pi^0$ channel with $T=0,\ \mu \neq 0$ and $k^2_3\neq0,\ \mathbf{k}_\perp^2= 0$. With $0<\omega<{\text {PB}}^{d,-}_L(0,0)$, $\Phi_{\pi^0}=\pi$, and with ${\text {PB}}^{d,-}_L(0,0)<\omega<{\text {PB}}^{d,+}_L(0,0)$, $\Phi_{\pi^0}$ decreases from $\pi$ to $0$ continuously. After that, $\Phi_{\pi^0}$ keeps zero, until a jump from $0$ to $\pi$ at $\omega=\omega_{\delta}$. $\Phi_{\pi^0}$ keeps $\pi$ with $\omega_{\delta}+0^+<\omega<\omega^d_U(0,0)-0^+$, and jumps from $\pi$ to $\phi_U<\pi/2$ at $\omega=\omega^d_U(0,0)+0^+$. For $\omega^d_U(0,0)+0^+<\omega<\omega^d_U(1,1)-0^+$, $\Phi_{\pi^0}$ changes nonmonotonically, which crosses $\pi/2$ at $\omega=\omega_{\text{BW}}$, reaches a local maximum ($<\pi$) at $\omega={\text {PB}}^{d,+}_U(0,0)$, and approaches $\pi$ at $\omega=\omega^d_U(1,1)-0^+$. At $\omega=\omega^d_U(1,1)+0^+$, a sudden jump from $\pi$ to $\phi_U<\pi/2$ is observed, and then $\Phi_{\pi^0}$ slightly decreases and then increases, which crosses $\Phi_{\pi^0}=\pi/2$ at $\omega=\omega_{\text{BW}}$ and reaches $\Phi_{\pi^0}=\pi$ at $\omega=\omega^d_U(2,2)-0^+$.

Figure~\ref{phase}(c2) magenta lines plot the $q\bar{q}$ scattering phase shift in $\pi^0$ channel with $T=0,\ \mu \neq 0$ and $k^2_3=0,\ \mathbf{k}_\perp^2\neq 0$. Scattering phase shift $\Phi_{\pi^0}$ shows a local maximum ($<\pi/2$) around ${\text {PB}}^{d,\pm}_L(1,0)$, and a local maximum ($>\pi/2$) with $\omega<\omega^d_L(1,0)-0^+$. At $\omega=\omega^d_L(1,0)+0^+$, $\Phi_{\pi^0}$ jumps to zero. After that, at $\omega_\delta$, $\Phi_{\pi^0}$ jumps from zero to $\pi$. It keeps $\pi$ until $\omega={\text {PB}}^{d,\pm}_L(2,0)$, and starts to decrease. With ${\text {PB}}^{d,\pm}_L(2,0)<\omega<\omega^d_U(0,1)-0^+$, $\Phi_{\pi^0}$ forms a dip structure, with a tiny platform around ${\text {PB}}^{d,\pm}_U(0,0)$. At $\omega=\omega^d_U(0,1)+0^+$, $\Phi_{\pi^0}$ jumps from $\pi$ to $\phi_U>\pi/2$, and then, it slightly increases and turns to decrease with increasing $\omega$. Since we meet too many unitary, Landau and PB thresholds, which make difficult to show the $q\bar{q}$ scattering phase shift in the $\pi^0$ channel in the whole $\omega$ region clearly. We just plot $\Phi_{\pi^0}$ with $0<\omega<0.6{\text{GeV}}$ to characterize its property.

\section{\label{sec:level4}Summary and outlook}

$q\bar{q}$ scattering phase shift in the $\pi^0$ channel $\Phi_{\pi^0}(\omega^2,\mathbf{k}_\perp^2,k^2_3)$ and ${\pi}^0$ meson spectral function $\rho_{\pi^0}(\omega^2,\mathbf{k}_\perp^2,k^2_3)$ under the external magnetic field $eB$ and finite meson momentum $\mathbf{k}_\perp^2,k^2_3$ are studied in the framework of a two-flavor Nambu-Jona-Lasinio (NJL) model. The $q\bar{q}$ scattering phase shift in the $\pi^0$ channel $\Phi_{\pi^0}$ is closely related to $\pi^0$ spectral function $\rho_{\pi^0}$. We consider three situations, chiral broken phase ($T=\mu=0$), chiral restoration phase ($T>T_{pc},\ \mu=0$) and chiral restoration phase ($T=0,\ \mu>\mu_{pc}$).

For $T=\mu=0$ and $T>T_{pc},\ \mu=0$ cases, ${\pi}^0$ meson spectral function $\rho_{\pi^0}$ shows a delta peak, several Breit-Wigner peaks and several non-Breit-Wigner peaks. The delta peak indicates the bound state of $\pi^0$ meson, and the Breit-Wigner peak means the resonant state of $\pi^0$ meson. For $T=0,\ \mu>\mu_{pc}$ case, Pauli-blocking effect plays a role, which changes the inner structure of these Breit-Wigner peaks and non-Breit-Wigner peaks. The multiple peak structure is caused by the external magnetic field. 

The $q\bar{q}$ scattering phase shift in the $\pi^0$ channel $\Phi_{\pi^0}$ shows a jump from $0$ to $\pi$ when $\pi^0$ meson is in bound state. When $\pi^0$ meson is in resonant state, $\Phi_{\pi^0}$ has the value $\pi/2$ and changes continuously. In large $\omega$ region, at the starting and end points of wide peaks of $\pi^0$ spectral function, $\Phi_{\pi^0}$ jumps abruptly (from $\pi$ to finite value or from finite value to $0$), and such jumps are caused by the external magnetic field. In the low $\omega$ region, $\Phi_{\pi^0}$ changes continuously.

Finite momentum $\mathbf{k}_\perp^2$ or $k^2_3$ modifies the spectral function $\rho_{\pi^0}$ and scattering phase shift $\Phi_{\pi^0}$, which demonstrates the anisotropy in the system, induced by the external magnetic field. The experimental observability of this anisotropy requires further investigation.

The results of $q\bar{q}$ scattering phase shift in the $\pi^0$ channel $\Phi_{\pi^0}(\omega^2,\mathbf{k}_\perp^2,k^2_3)$ and ${\pi}^0$ meson spectral function $\rho_{\pi^0}(\omega^2,\mathbf{k}_\perp^2,k^2_3)$ can provide input for understanding the thermodynamic properties in the magnetized quark-meson system. Usually, we start from a Lagrangian with current quarks and two-body interaction in scalar and pseudo-scalar channels. In mean-field approximation, the corresponding thermodynamic potential is that of a quark gas of constituent mass $m_q$. Beyond mean-field level, the interaction between quarks and antiquarks appears in the form of meson state and scattering phase shift. This method has been considered in the case without electromagnetic fields, which is more general than its application in NJL model, and resembles the Beth-Uhlenbeck formula for the second virial coefficient for relativistic particles~\cite{HUFNER1994225,ZHUANG1994525,BLASCHKE2014228}. We will generalize the method to the case with electromagnetic fields in the future, which may be helpful to understand the properties of QCD matter, such as fluctuations and correlations~\cite{PhysRevD.86.034509,PhysRevD.88.014009,Borsanyi2018,fm5f-ctr3}, and QCD phase transition~\cite{Bali2012,PhysRevD.83.034016,PhysRevD.98.054509,Endrodi2015}, such as inverse magnetic catalysis phenomena~\cite{PhysRevD.86.071502,Bali2013,PhysRevLett.110.031601,PhysRevD.94.036007,PhysRevD.91.116010}.\\

\noindent {\bf Acknowledgement:} Shijun Mao is supported by the National Natural Science Foundation of China under Grant No.12275204. Guoyun Shao is supported by the National Natural Science Foundation of China under Grant No.12475145 and Natural Science Basic Research Plan in Shaanxi Province of China (Program No. 2024JC-YBMS-018).\\

\appendix
\section{$eB=0$}
\begin{widetext}
Taking the zero magnetic field limit, we recover the $\pi^0$ polarization function in~\cite{RevModPhys.64.649,ZHUANG1994525}:
\begin{equation}
	\label{Repolar}
	\begin{aligned}
		&\Pi^{{\cal {R}}e}(\omega^2,\vec{k}^2)=\operatorname{Re}[\Pi_{\pi^0}((\omega+i\epsilon)^2,\vec{k}^2)]\\
		& =-4N_c\int\frac{d^3{\vec{q}}}{(2\pi)^3}\left\{\frac{F(E_{\vec{q}}-\mu)-F(-E_{\vec{q}}-\mu)}{2E_{\vec{q}}}+\frac{F(E_{\vec{q}+\vec{k}}-\mu)-F(-E_{\vec{q}+\vec{k}}-\mu)}{2E_{\vec{q}+\vec{k}}}\right.\\
		&\quad \quad \quad \quad \quad \quad \quad \quad \quad -\frac{\omega^2-\vec{k}^2}{4E_{\vec{q}+\vec{k}}E_{\vec{q}}}\left(\frac{F(E_{\vec{q}}-\mu)-F(E_{\vec{q}+\vec{k}}-\mu)}{\omega+E_{\vec{q}}-E_{\vec{q}+\vec{k}}}+ \frac{F(-E_{\vec{q}}-\mu)-F(-E_{\vec{q}+\vec{k}}-\mu)}{\omega-E_{\vec{q}}+E_{\vec{q}+\vec{k}}}\right.\\
		&\quad \quad \quad \quad \quad \quad \quad \quad \quad \quad  \quad \quad \quad \quad \quad \left.\left.-\frac{F(-E_{\vec{q}}-\mu)-F(E_{\vec{q}+\vec{k}}-\mu)}{\omega-E_{\vec{q}}-E_{\vec{q}+\vec{k}}}-\frac{F(E_{\vec{q}}-\mu)-F(-E_{\vec{q}+\vec{k}}-\mu)}{\omega+E_{\vec{q}}+E_{\vec{q}+\vec{k}}}\right)\right\}\\
	\end{aligned}
\end{equation}
and
\begin{equation}
	\label{Impolar}
	\begin{aligned}
		&\Pi^{{\cal {I}}m}(\omega^2,\vec{k}^2)=\operatorname{Im}[\Pi_{\pi^0}((\omega+i\epsilon)^2,\vec{k}^2)]=-2\pi N_c\int\frac{d^3{\vec{q}}}{(2\pi)^3}\frac{\omega^2-\vec{k}^2}{2E_{\vec{q}+\vec{k}}E_{\vec{q}}}\\
		&\times\left(\left[F(E_{\vec{q}}-\mu)-F(E_{\vec{q}+\vec{k}}-\mu)\right]\delta(\omega+E_{\vec{q}}-E_{\vec{q}+\vec{k}})+\left[F(-E_{\vec{q}}-\mu)-F(-E_{\vec{q}+\vec{k}}-\mu)\right]\delta(\omega-E_{\vec{q}}+E_{\vec{q}+\vec{k}})\right.\\
		&\quad \quad \left.-\left[F(E_{\vec{q}}-\mu)-F(-E_{\vec{q}+\vec{k}}-\mu)\right]\delta(\omega+E_{\vec{q}}+E_{\vec{q}+\vec{k}})-\left[F(-E_{\vec{q}}-\mu)-F(E_{\vec{q}+\vec{k}}-\mu)\right]\delta(\omega-E_{\vec{q}}-E_{\vec{q}+\vec{k}})\right)\\
	\end{aligned}
\end{equation}
with quark energy $E_{\vec{q}}=\sqrt{\vec{q}^2+m_q^2}$ and $E_{\vec{q}+\vec{k}}=\sqrt{(\vec{q}+\vec{k})^2+m_q^2}$. Here, the quark mass $m_q$ is determined by the gap equation $m_0-(1-2G{\cal J}_1)m_q=0$, with ${\cal J}_1=4N_c\int\frac{d^3\vec{q}}{(2\pi)^3}\frac{1-F(E_{\vec{q}}-\mu)-F(E_{\vec{q}}+\mu)}{E_{\vec{q}}}$.

In this case, we have one unitary threshold
\begin{equation}
	\label{B0thu}
	\omega_{\text{th},U}=2\sqrt{m_q^2+\vec{k}^2/4},
\end{equation}
and one Landau threshold
\begin{equation}
	\label{B0thl}
	\omega_{\text{th},L}=|\vec{k}|.
\end{equation}
Different from the case with finite magnetic field $eB \neq 0$, no divergence happens in these thresholds (See Table~\ref{tab:threshold_eB0}).
\begin{table}[htbp]
	\centering
	\renewcommand{\arraystretch}{1.3}
	\caption{Behavior near different thresholds with $eB=0$ for $\vec{k}^2=0$ and $\vec{k}^2 \neq 0$.}
	\label{tab:threshold_eB0}
	
	\begin{tabular}{ccccc}
		\hline
		&
		$\Pi^{{\cal {R}}e}(\omega^2,\vec{k}^2)$ &
		$\Pi^{{\cal {I}}m}(\omega^2,\vec{k}^2)$ &
		$\Phi_{\pi^0}(\omega^2,\vec{k}^2)$ &
		$\rho_{\pi^0}(\omega^2,\vec{k}^2)$
		\\
		\hline
		
		$\omega \to(\omega_{\text{th},U})^-$ &
		finite &
		0 &
		$0/\pi$ (${\text {for}}$ $\operatorname{sgn}[1-2G\Pi^{{\cal {R}}e}]=\pm$) &
		0 \\[2mm]
		
		$\omega \to(\omega_{\text{th},U})^+$ &
		finite &
		$0$ &
		$ 0/\pi$ (${\text {for}}$ $\operatorname{sgn}[1-2G\Pi^{{\cal {R}}e}]=\pm$) &
		$0$ \\[2mm]
		
		$\omega \to(\omega_{\text{th},L})^-$ &
		finite &
		$0$ &
		$ 0/\pi$ (${\text {for}}$ $\operatorname{sgn}[1-2G\Pi^{{\cal {R}}e}]=\pm$) &
		$0$ \\[2mm]
		
		$\omega \to(\omega_{\text{th},L})^+$ &
		finite &
		0 &
		$0/\pi$ (${\text {for}}$ $\operatorname{sgn}[1-2G\Pi^{{\cal {R}}e}]=\pm$) &
		0 \\
		
		\hline
	\end{tabular}
\end{table}

Moreover, in the zero temperature and finite quark chemical potential case, four PB thresholds are
\begin{equation}
	\text{PB}^{\pm}_{U}=\sqrt{\mu^2+\vec{k}^2\pm2|\vec{k}|\sqrt{\mu^2-m_q^2}}+\mu,
\end{equation}
and
\begin{equation}
	\text{PB}^{\pm}_{L}=\sqrt{\mu^2+\vec{k}^2\pm2|\vec{k}|\sqrt{\mu^2-m_q^2}}-\mu.
\end{equation}
Note that only PB thresholds with positive value play the role in the calculation.

\end{widetext}

%%%%%%%%%%%%%%%%%%%%%%%%%%%%%%%%%%%%%%%%%%%%%%%%%%%%%%%%%%%%%%%
\begin{figure}[htbp]
	\centering
	\includegraphics[width=0.48\textwidth]{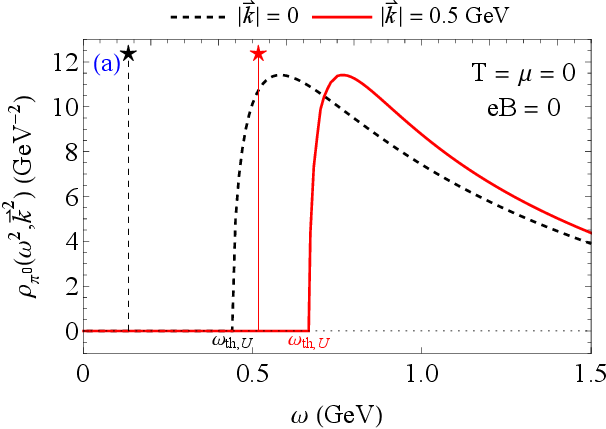}
	\vspace{2pt}
	\includegraphics[width=0.48\textwidth]{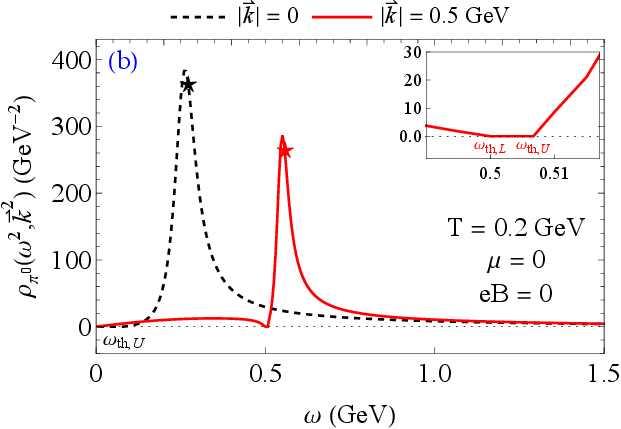}
	\vspace{2pt}
	\includegraphics[width=0.48\textwidth]{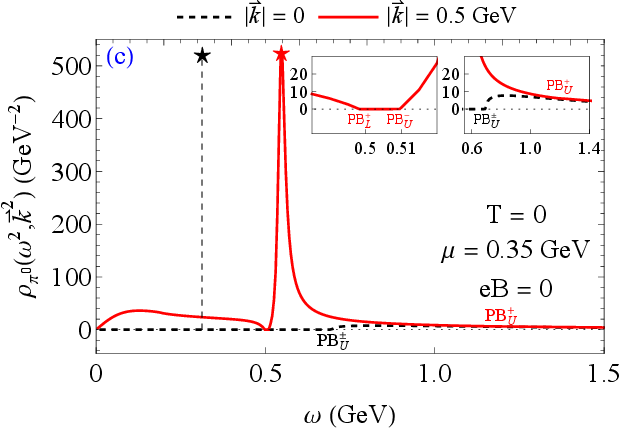}
	\centering
\caption{Spectral function of $\pi^0$ meson is shown as a function of $\omega$ for different $T$ and $\mu$ at $eB=0$. The black dashed and red solid curves represent the case with zero momentum $|\vec{k}|=0$ and with finite momentum $|\vec{k}|=0.5$ GeV, respectively. Here $\omega_{th,U/L}$ denotes unitary and Landau thresholds, and ${\text{PB}}_{U/L}^{\pm}$ denotes Pauli-blocking thresholds. The star marks the location of the solutions of pole equation $1-2G\Pi^{{\cal {R}}e}(\omega^2,\vec{k}^2)=0$.}
	\label{spectraleb0}
\end{figure}
%%%%%%%%%%%%%%%%%%%%%%%%%%%%%%%%%%%%%%%%%%%%%%%%%%%%%%%%%%%%%%%

Figure~\ref{spectraleb0} plots the $\pi^0$ spectral function at $eB=0$. We focus on the effect of finite momentum ($\vec{k}^2 \neq 0$) in chiral symmetry broken phase ($T=\mu=0$) and in chiral symmetry restored phase ($T>T_{pc}$ or $\mu >\mu_{pc}$). Here, the solution of pole equation $1-2G\Pi^{{\cal {R}}e}(\omega^2,\vec{k}^2)=0$ is marked by the star.

Figure~\ref{spectraleb0}(a) black line (red line) depicts the $\pi^0$ spectral function in chiral symmetry broken phase with $T=\mu=0$ and vanishing momentum $\vec{k}^2 = 0$ (finite momentum $\vec{k}^2 \neq 0$). They show the typical results of spectral function, with a delta peak and a continuum part. The location of the delta peak is determined by the pole equation $1-2G\Pi^{{\cal {R}}e}(\omega^2,\vec{k}^2)=0$. The continuum part begins to appear at unitary threshold $\omega=\omega_{\text{th},U}$. Due to the Lorentz symmetry, the black and red lines coincide with each other by the replacement $\omega \rightarrow \sqrt{\omega^2+{\vec k}^2}$.

Figure~\ref{spectraleb0}(b) black line (red line) depicts the $\pi^0$ spectral function in chiral symmetry restored phase with $T\neq 0,\ \mu=0$ and vanishing momentum $\vec{k}^2 = 0$ (finite momentum $\vec{k}^2 \neq 0$). With $\vec{k}^2 = 0$, the non-cross terms and unitary threshold control the spectral function, and the cross terms and the Landau threshold do not contribute to spectral function. It has a single Breit-Wigner peak, which begins at unitary threshold $\omega=\omega_{\text{th},U}$ and has a long tail extending to large value of $\omega$. With finite momentum $\vec{k}^2 \neq 0$, the cross terms and the Landau threshold also make contributions to spectral function. In low $\omega$ region, we observe the continuum part with $0<\omega<\omega_{\text{th},L}$. A Breit-Wigner peak starts at $\omega=\omega_{\text{th},U}$, and has a long tail extending to large value of $\omega$. In this figure, the summit of Breit-Wigner peak is slightly deviated from the star, due to the non-zero imaginary part of polarization function in Eq.\eqref{Impolar}.

Figure~\ref{spectraleb0}(c) black line (red line) depicts the $\pi^0$ spectral function in chiral symmetry restored phase with $T=0,\ \mu\neq 0$ and vanishing momentum $\vec{k}^2 = 0$ (finite momentum $\vec{k}^2 \neq 0$). In this case, the Pauli-blocking effect plays the role in spectral function. With $\vec{k}^2 = 0$, we have one positive PB threshold $\text{PB}_{U}^\pm=2\mu$. The spectral function shows a delta peak, marked by the star, and a continuum part, starting at PB threshold $\omega=\text{PB}_{U}^\pm=2\mu$. With $\vec{k}^2 \neq 0$, we have three positive PB thresholds $0<\text{PB}_{L}^+ <\text{PB}_{U}^- <\text{PB}_{U}^+$. The spectral function shows a continuum part with $0<\omega<\text{PB}_{L}^+ $ and a Breit-Wigner peak, which begins at $\omega=\text{PB}_{U}^-$ and has a long tail extending to large value of $\omega$.

%%%%%%%%%%%%%%%%%%%%%%%%%%%%%%%%%%%%%%%%%%%%%%%%%%%%%%%%%%%%%%%%
\begin{figure}[htbp]
	\centering
	\includegraphics[width=0.5\textwidth]{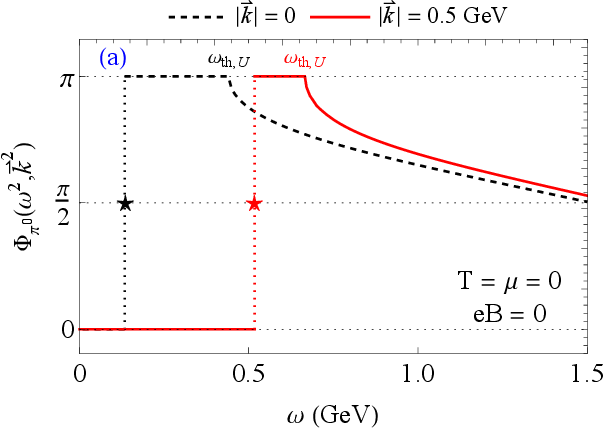}
	\vspace{2pt}
	\includegraphics[width=0.5\textwidth]{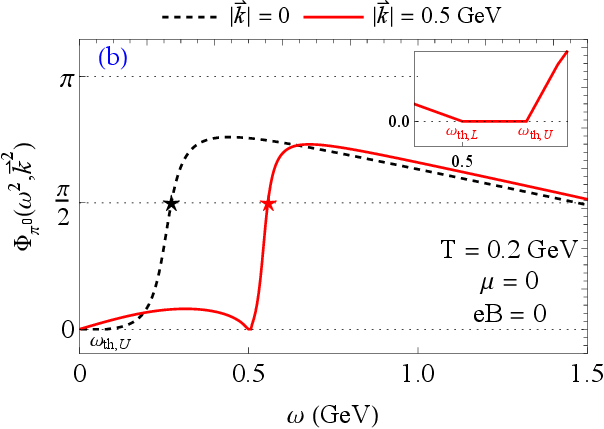}
	\vspace{2pt}
	\includegraphics[width=0.5\textwidth]{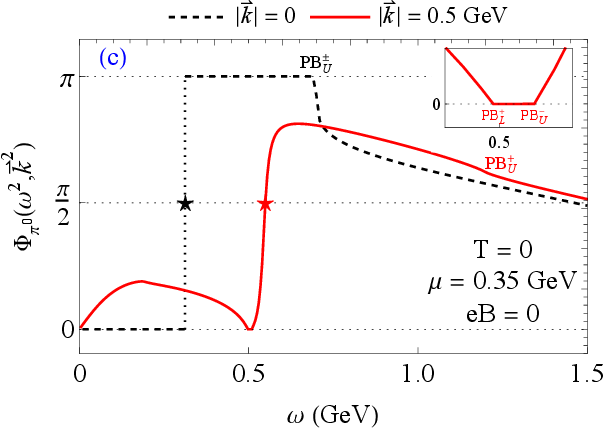}
	\caption{The $q\bar{q}$ scattering phase shift in the $\pi^0$ channel $\Phi_{\pi^0}$ is shown as a function of $\omega$ for different $T$ and $\mu$ at $eB=0$. The black dashed and red solid curves represent the case with zero momentum $|\vec{k}|=0$ and with finite momentum $|\vec{k}|=0.5$ GeV, respectively. Here $\omega_{th,U/L}$ denotes unitary and Landau thresholds, and ${\text{PB}}_{U/L}^{\pm}$ denotes Pauli-blocking thresholds. The star marks the location of the solutions $\Phi_{\pi^0}(\omega^2,\vec{k}^2)=\pi/2$.}
	\label{phaseeb0}
\end{figure}
%%%%%%%%%%%%%%%%%%%%%%%%%%%%%%%%%%%%%%%%%%%%%%%%%%%%%%%%%%%%%%%%

Figure~\ref{phaseeb0} plots the $q\bar{q}$ scattering phase shift in $\pi^0$ channel $\Phi_{\pi^0}$ with vanishing magnetic field $(eB=0)$.  We focus on the effect of finite momentum ($\vec{k}^2 \neq 0$) in chiral symmetry broken phase ($T=\mu=0$) and in chiral symmetry restored phase ($T>T_{pc}$ or $\mu >\mu_{pc}$).

According to the definition of $q\bar{q}$ scattering phase shift in Eq.\eqref{phasedefinition}, when we meet the pole of $\pi^0$ propagator $1 - 2G\Pi^{{\cal {R}}e}(\omega^2_{\delta},{\vec k}^2)=0$ with vanishing imaginary part of polarization function $\Pi^{{\cal {I}}m}(\omega^2_{\delta},{\vec k}^2)=0$, we obtain $\Phi_{\pi^0}(\omega^2,{\vec k}^2)=\pi\Theta(\omega^2-\omega_{\delta}^2)$ with $\omega$ near $\omega_{\delta}$. At the pole of $\pi^0$ propagator $1 - 2G\Pi^{{\cal {R}}e}(\omega^2_{\text{BW}},{\vec k}^2)=0$ with nonvanishing imaginary part of polarization function $\Pi^{{\cal {I}}m}(\omega^2_{\text{BW}},{\vec k}^2)\neq 0$, we have $\Phi_{\pi^0}(\omega^2_{\text{BW}},{\vec k}^2)=\pi/2$. Therefore, we mark a star at $\Phi_{\pi^0}=\pi/2$ in all panels of Figure~\ref{phaseeb0}.

Figure~\ref{phaseeb0}(a) black line plots the $q\bar{q}$ scattering phase shift in $\pi^0$ channel $\Phi_{\pi^0}$ in chiral broken phase with $T=\mu=0$ and $\vec{k}^2 = 0$. $\Phi_{\pi^0}$ keeps zero, until the jump from $0$ to $\pi$ at $\omega=\omega_\delta$. Further increasing $\omega$, $\Phi_{\pi^0}$ keeps $\pi$, and starts to decrease at the unitary threshold $\omega=\omega_{\text{th,U}}$. Figure~\ref{phaseeb0}(a) red line plots the $q\bar{q}$ scattering phase shift in $\pi^0$ channel $\Phi_{\pi^0}$ with $T=\mu=0$ and $\vec{k}^2 \neq 0$, which recovers the results in black line by a shift $\omega \rightarrow \sqrt{\omega^2-{\vec k}^2}$, due to the Lorentz symmetry.

Figure~\ref{phaseeb0}(b) black line plots the $q\bar{q}$ scattering phase shift in $\pi^0$ channel $\Phi_{\pi^0}$ in chiral restoration phase with $T\neq0, \ \mu=0$ and $\vec{k}^2 = 0$.  $\Phi_{\pi^0}$ keeps zero until unitary threshold $\omega=\omega_{\text{th,U}}$. It starts to increase continuously, which crosses $\pi/2$ at $\omega=\omega_{\text{BW}}$. After reaching a local maximum ($<\pi$), $\Phi_{\pi^0}$ turns to decrease with increasing $\omega$. Figure~\ref{phaseeb0}(b) red line plots the $q\bar{q}$ scattering phase shift in $\pi^0$ channel $\Phi_{\pi^0}$ in chiral restoration phase with $T\neq0, \ \mu=0$ and $\vec{k}^2 \neq 0$. In addition to the unitary threshold $\omega_{\text{th,U}}$, we meet the Landau threshold $\omega_{\text{th,L}}$, which has a lower value than $\omega_{\text{th,U}}$. In low $\omega$ region, $\Phi_{\pi^0}$ increases from zero and then decreases back to zero at Landau threshold $\omega=\omega_{\text{th,L}}$. Further increasing $\omega$, $\Phi_{\pi^0}$ in red line is similar to that in black line.

Figure~\ref{phaseeb0}(c) black line plots the $q\bar{q}$ scattering phase shift in $\pi^0$ channel $\Phi_{\pi^0}$ in chiral restoration phase with $T=0, \ \mu\neq 0$ and $\vec{k}^2 = 0$. $\Phi_{\pi^0}$ shows similar behavior to that in Figure~\ref{phaseeb0}(a) black line. There are two differences. On the one hand, the value of $\omega_\delta$ is modified by the quark chemical potential. On the other hand, due to the Pauli blocking effect, the decreasing behavior starts at the PB threshold $\omega=\text{PB}_{U}^\pm=2\mu$, not at unitary threshold. Figure~\ref{phaseeb0}(c) red line plots the $q\bar{q}$ scattering phase shift in $\pi^0$ channel $\Phi_{\pi^0}$ in chiral restoration phase with $T=0, \ \mu\neq 0$ and $\vec{k}^2 \neq 0$. $\Phi_{\pi^0}$ looks similar to that in Figure~\ref{phaseeb0}(b) red line, with the replacement of $\omega_{\text{th,L}} \rightarrow \text{PB}_{L}^+$ and $\omega_{\text{th,U}} \rightarrow \text{PB}_{U}^-$ due to the Pauli blocking effect.

\bibliography{refs}

\end{document}